\documentclass{article}

\usepackage[preprint, nonatbib]{neurips_2025}

\usepackage{float}
\usepackage{wrapfig}
\usepackage{listings}
\usepackage[utf8]{inputenc} 
\usepackage[T1]{fontenc}    
\usepackage[backref=page]{hyperref} 
\usepackage{url}            
\usepackage{booktabs}       
\usepackage{amsfonts}       
\usepackage{nicefrac}       
\usepackage{microtype}      
\usepackage{xcolor}         
\usepackage{graphicx}
\usepackage{amsmath}
\usepackage[english]{babel}
\usepackage{enumitem}

\usepackage{xcolor}

\colorlet{punct}{red!60!black}
\definecolor{background}{HTML}{EEEEEE}
\definecolor{delim}{RGB}{20,105,176}
\colorlet{numb}{magenta!60!black}

\lstdefinelanguage{json}{
    basicstyle=\ttfamily\small,
    showstringspaces=false,
    breaklines=true,
    backgroundcolor=\color{white},
    frame=single,        
    numbers=none,      
    numbersep=5pt,
    numberstyle=\tiny\color{gray},
    literate=
     *{0}{{{\color{numb}0}}}{1}
      {1}{{{\color{numb}1}}}{1}
      {2}{{{\color{numb}2}}}{1}
      {3}{{{\color{numb}3}}}{1}
      {4}{{{\color{numb}4}}}{1}
      {5}{{{\color{numb}5}}}{1}
      {6}{{{\color{numb}6}}}{1}
      {7}{{{\color{numb}7}}}{1}
      {8}{{{\color{numb}8}}}{1}
      {9}{{{\color{numb}9}}}{1}
      {:}{{{\color{punct}{:}}}}{1}
      {,}{{{\color{punct}{,}}}}{1}
      {\{}{{{\color{delim}{\{}}}}{1}
      {\}}{{{\color{delim}{\}}}}}{1}
      {[}{{{\color{delim}{[}}}}{1}
      {]}{{{\color{delim}{]}}}}{1},
}

\lstdefinelanguage{yaml}{
  keywords={true,false,null,y,n},
  keywordstyle=\color{blue}\bfseries,
  basicstyle=\ttfamily\small,
  sensitive=false,
  comment=[l]{\#},
  commentstyle=\color{gray}\itshape,
  stringstyle=\color{green!40!black},
  morestring=[b]',
  morestring=[b]"',
  stringstyle=\color{purple},
  morecomment=[s]{<!--}{-->},
  identifierstyle=\color{black},
  numberstyle=\color{gray},
  showstringspaces=false,
  morekeywords={version, services, build, image, container_name, volumes, ports, environment, depends_on, networks},
  literate=
    {>}{{\textcolor{orange}{>}}}1
    {|}{{\textcolor{orange}{|}}}1
    {\ -\ }{{\textcolor{orange}{\ -\ }}}3,
}

\graphicspath{ {./figures/} }

\title{Cybernaut: Towards Reliable Web Automation}

\author{%
    Ankur Tomar\thanks{Applied AI, Amazon.com, Bellevue, WA, USA}  \\
    \texttt{axtomar@amazon.com} \\
    \And
    Hengyue Liang$^*$ \\
    \texttt{hengyue@amazon.com} \\
    \And
    Indranil Bhattacharya$^*$ \\
    \texttt{bindrani@amazon.com} \\
    \And
    Natalia Larios$^*$ \\
    \texttt{natalild@amazon.com} \\
    \And
    Francesco Carbone$^*$ \\
    \texttt{carbonef@amazon.lu}
}

\begin{document}

\maketitle

\begin{abstract} The emergence of AI-driven web automation through Large Language Models (LLMs) offers unprecedented opportunities for optimizing digital workflows. However, deploying such systems within industry's real-world environments presents four core challenges: (1) ensuring consistent execution, (2) accurately identifying critical HTML elements, (3) meeting human-like accuracy in order to automate operations at scale and (4) the lack of comprehensive benchmarking data on internal web applications. Existing solutions are primarily tailored for well-designed, consumer-facing websites (e.g., Amazon.com, Apple.com) and fall short in addressing the complexity of poorly-designed internal web interfaces. To address these limitations, we present \emph{Cybernaut}, a novel framework to ensure high execution consistency in web automation agents designed for robust enterprise use. Our contributions are threefold: (1) a Standard Operating Procedure (SOP) generator that converts user demonstrations into reliable automation instructions for linear browsing tasks, (2) a high-precision HTML DOM element recognition system tailored for the challenge of complex web interfaces, and (3) a quantitative metric to assess execution consistency. The empirical evaluation on our internal benchmark demonstrates that using our framework enables a 23.2\% improvement (from 72\% to 88.68\%) in task execution success rate over the baseline~\cite{browser_use2024}. Cybernaut identifies consistent execution patterns with 84.7\% accuracy, enabling reliable confidence assessment and adaptive guidance during task execution in real-world systems. These results highlight Cybernaut's effectiveness in enterprise-scale web automation and lay a foundation for future advancements in web automation.
\end{abstract}

\section{Introduction}
\label{sec:introduction}
\begin{figure*}[!htbp]
\centering
\includegraphics[width=\linewidth]{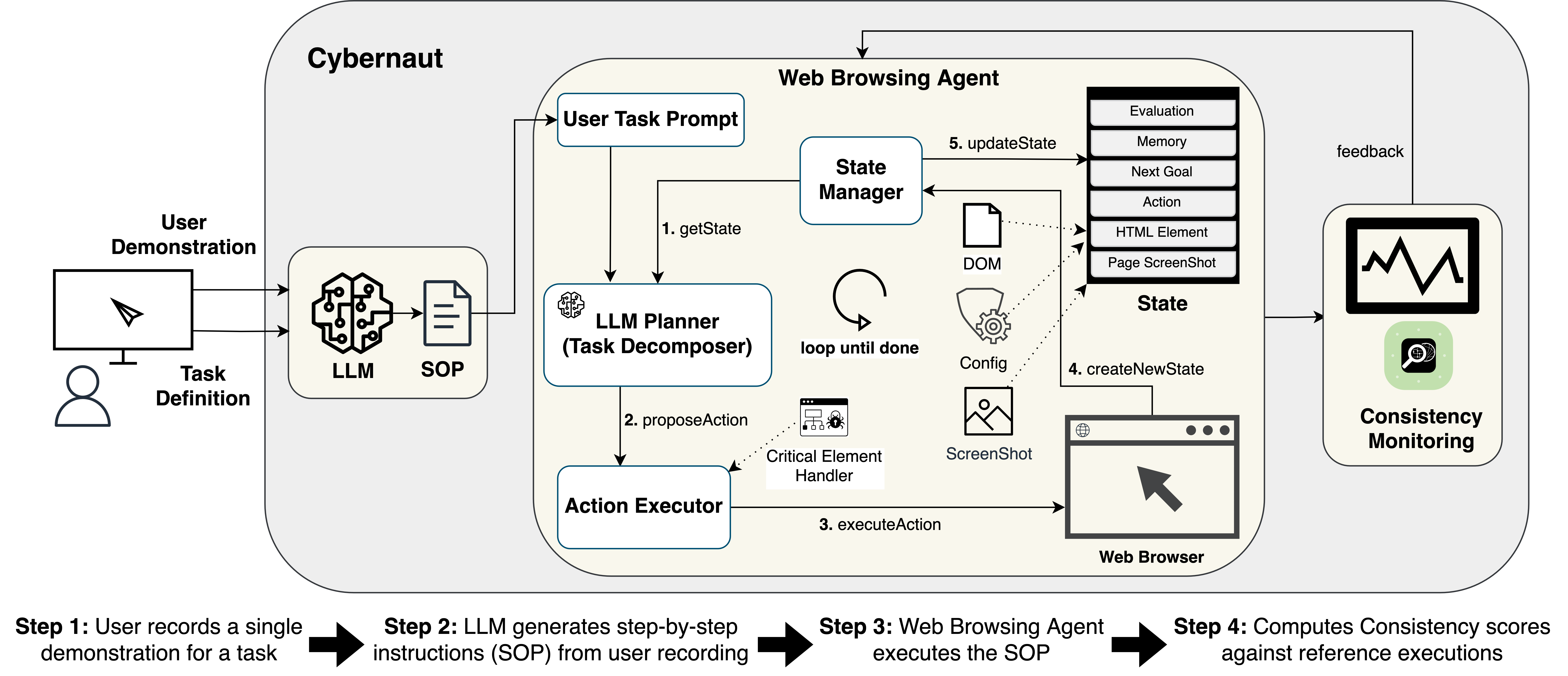}
\caption{Architectural overview of Cybernaut's workflow pipeline, illustrating the transformation of user demonstrations into executable SOPs through LLM processing, web agent execution, and consistency monitoring feedback loops.}
\label{fig:cybernaut}
\end{figure*}
Recent advances in Large Language Models (LLMs) have provided them with remarkable reasoning capabilities that enable the development of autonomous agents. Autonomous agents can automate and optimize Knowledge Operations, which are repetitive and well defined tasks involving the creation, manipulation,  or classification of knowledge. A common step in Knowledge Operation tasks consists of interacting with a web application to retrieve, submit and manipulate data. Instruction-based web browsing tools enable automation for these types of tasks. They represents a significant frontier in this domain \cite{ning2025survey}, where LLMs are equipped with browsing tools (e.g., navigation, clicking, scrolling, form filling, etc.) to interpret website structures and execute required actions. While several solutions have emerged, including Operator~\cite{openai2025cua}, Computer use~\cite{anthropic2024computeruse} and Browser-Use~\cite{browser_use2024}, their evaluation methodologies often rely on selecting optimal performances from multiple attempts, highlighting a critical limitation in consistency and reliability. Moreover, public benchmarks like WebVoyager~\cite{he-etal-2024-webvoyager} may not effectively represent scenarios where agent interacts with a limited set of websites but demand high consistency and quality. This disparity between benchmark performance and practical requirements underscores the need for more focused evaluation methodologies.

Enterprise web automation faces several challenges particular to real-world scenarios, especially when performing repetitive tasks with dynamic parameters — for example, retrieving hazardous material ratings for different ASINs. Existing solutions often depend on brittle, hard-coded, element-based approaches that are highly sensitive to UI changes. Also, accurately detecting \emph{interactable} elements on web pages remains a significant challenge. While tools like \emph{Selenium} and \emph{Playwright} offer mechanisms to generate summarized snapshots of HTML elements, the heterogeneity of web interfaces often leads to detection failures, resulting in incomplete action spaces and reduced task accuracy. Furthermore, existing browsing agents rely heavily on detailed task descriptions from users. While comprehensive instructions can improve task specificity, they often result in brittle solutions that fail when websites undergo minor changes.

To address these limitations, we introduce \emph{Cybernaut}, a framework that provides a robust mechanism for measuring and ensuring high consistency in repeated executions of web automation agents, while simultaneously meeting the stringent accuracy requirements. Cybernaut is built on the principles of demonstration-based learning \cite{correia2023survey}, with a particular emphasis on the challenges encountered in real-world environments. Our solution addresses these limitations through three key innovations: (1) automated generation of high-level execution steps from user demonstrations, (2) robust element detection and interaction handling, and (3) quantitative consistency evaluation across multiple executions. Figure~\ref{fig:cybernaut} illustrates the architectural overview of Cybernaut's workflow pipeline.

\section{Related work}
\label{sec:related_work}
\subsection{Web Navigation Agents}
\label{subsec:web_agents}
Recent advances in LLMs have led to diverse approaches in web agent design, categorized by their user interaction paradigms, including personalization, multi-modality, and grounding mechanisms. These range from conversational agents like WebLINX \cite{lu2024_weblinx} that frame web navigation as a dialogue task, to API-centric approaches \cite{song2025_api_browsingapi} offering structured interfaces between LLMs and web environments. Visual grounding has emerged as a promising direction, with studies like \cite{zheng2024_gpt4_webagent} demonstrating GPT-4V's effectiveness in grounding actions through element highlighting. Nova-Act~\cite{nova_act} achieve best-in-class performance on benchmarks like ScreenSpot and GroundUI Web which most directly measure the ability of model to actuate the web. Our work builds upon these approaches, particularly focusing on grounding-aided interactive element detection. We introduce novel methods for converting single demonstrations into robust steps that the agent can follow consistently and measure the execution's consistency. This advancement addresses the critical need for consistency and reliability, pushing the field forward in terms of practical applicability and performance.

\textbf{Browser-Use Framework:} Although our framework is agnostic to the underlying web agent, for the purpose of this paper, we use Browser-Use \cite{browser_use2024}, as it is an open-source project that has gained significant traction for web automation. It provides a sophisticated infrastructure for creating AI agents capable of autonomous web navigation, interaction, and information extraction. Our work extends the framework by effectively bridging the gap between user intent and consistent web agent execution.

\textbf{Perception of Interactive Elements:} Graphical User Interface (GUI) perception in LLM-based agents follows two main approaches. Single-modal LLMs \cite{wen2024a_autodroid_automation, li2020_mappinglang_instructions} use separate modules for GUI processing, while multi-modal LLMs \cite{hong2024cogagentvisuallanguagemodel, wang2024_mobileagent_multimod} handle visual and textual information in an integrated manner. Recent advancements in UI understanding include Ferret-UI's \cite{you2024_ferretui_ground} "any-resolution" approach, OmniParser's \cite{lu2024_omniparser_vision} vision-based parsing, and Iris's \cite{ge2025_irisb_guicomplex} enhanced processing of heterogeneous GUI information, addressing key challenges in annotation bias, modality misalignment, and architectural limitations.

\subsection{Benchmarks for Browser and Computer Agents}
\label{subsec:benchmark_browser_computer_use_agents}

\textbf{Web Navigation Benchmarks and Environments:} Several benchmarks have been proposed to evaluate LLM-based web agents. WebVoyager \cite{he-etal-2024-webvoyager} focuses on end-to-end web agents navigating real-world websites, and introduces advanced DOM element recognition techniques. VisualWebArena \cite{koh2024_visualwebarena} extends this by incorporating multi-modal agents and visual grounding for web tasks. Mind2Web \cite{deng2024_mid2web_neurips} provides a large-scale dataset spanning 137 websites and 31 domains, emphasizing generalist web agents. While these benchmarks primarily evaluate the successful completion of online tasks, some - like Mind2Web - also measure execution trace alignment with human demonstrations.

\textbf{Mobile and Cross-Platform Environments:} AndroidWorld \cite{rawles2025_androidworld} presents a comprehensive benchmark for mobile interfaces, featuring 116 programmatic tasks across 20 real-world Android apps. This environment enables testing under varying conditions through parameterized task generation. The CRAB benchmark \cite{xu2024_crabxenv_bench} extends this to cross-platform scenarios, encompassing both desktop and mobile environments, and encouraging development of generalizable agents.

\textbf{Execution-Based Evaluation Approaches:} Existing approaches differ in how they evaluate agents. Benchmarks like WebVoyager and VisualWebArena focus on end-to-end task completion in live environments, while Mind2Web and AndroidWorld emphasize execution trace-based evaluation against ground-truth demonstrations. The step verification approach (STEVE) by \cite{lu_steve_stepverif_2025} uses screenshots at every step to validate each action. CRAB \cite{xu2024_crabxenv_bench} proposes a graph-based evaluation of the agent trajectory in addition to traditional goal-based evaluations. Our work aligns with approaches for execution-based evaluation by introducing novel metrics measuring automation consistency and reliability in real-world settings.

\begin{figure*}[!htpb]
\centering
\includegraphics[width=\linewidth]{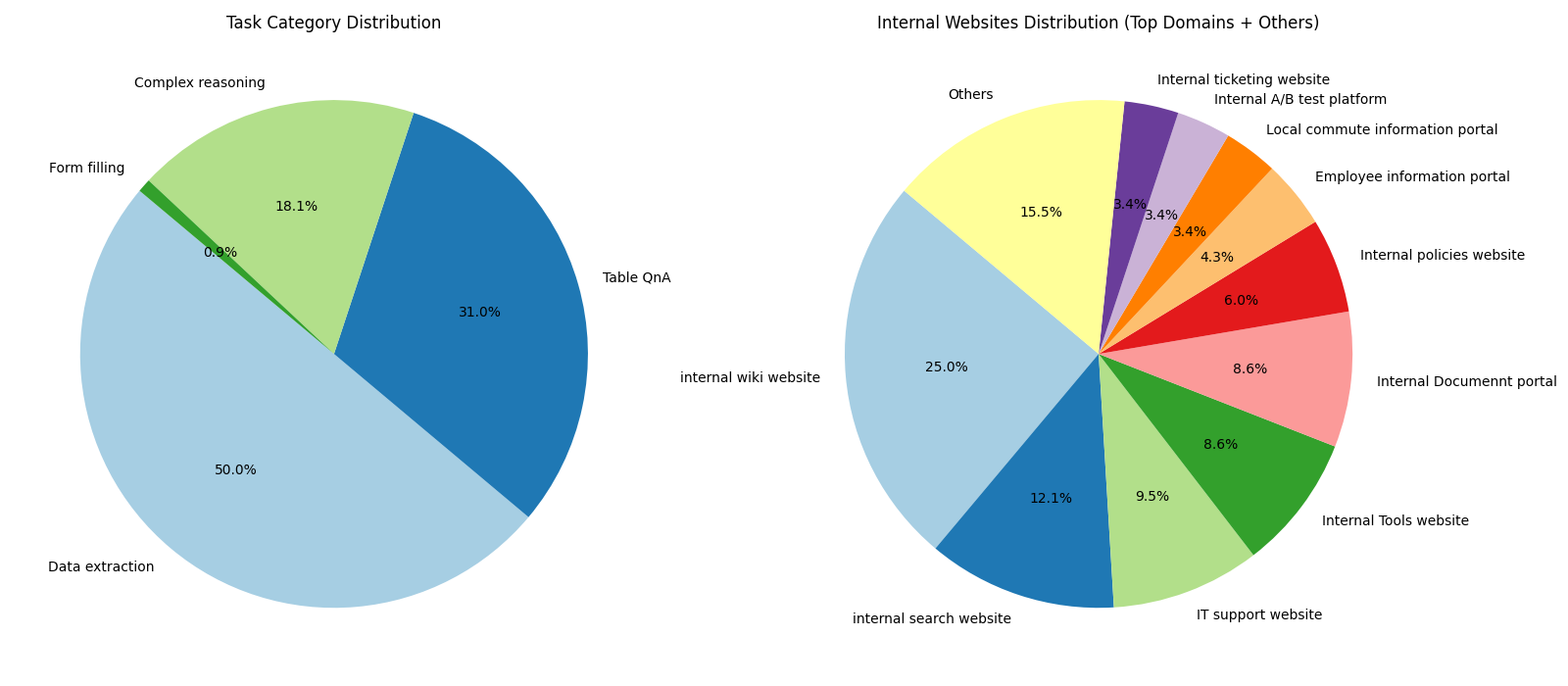}
\caption{Distribution of tasks across categories and internal website types in the benchmarking
dataset. The left pie chart shows the proportion of tasks across four categories: Data extraction, Table
QnA, Complex reasoning, and Form filling. The right pie chart highlights the most frequent internal
domains accessed, and aggregating low-frequency domains under "Others"}
\label{fig:dataset}
\end{figure*}

\section{Proposed Methodology}
\label{sec:proposed_method}
\subsection{Internal Web Benchmarking Data}
\label{subsec:internal_benchmark}
We introduce an internal benchmarking dataset to evaluate agent performance on our company's internal websites, which are primarily designed for internal operational needs over optimal HTML structure. This dataset enables other teams working on similar tasks to conduct standardized evaluations and make fair comparisons across different agent implementations. Our benchmarking dataset was constructed based on the following principles: selection of universally accessible internal websites with stable and consistent content; inclusion of historically stable tasks to ensure reproducibility; formulation of questions with unambiguous answers directly derivable from the web page; and coverage of diverse interaction types such as form-filling, summarization, table-based querying, and logical reasoning. The dataset consists of 117 tasks distributed across 25 internal domains, providing a representative testbed for evaluating enterprise-grade web automation agents in a zero-shot setting. A detailed breakdown of the dataset, including task category and domain distribution, is illustrated in Figure~\ref{fig:dataset}.

A majority of task interactions in the internal dataset are concentrated within high-frequency domains, which support core functions including policy access, IT support, and operational dashboards. Mid-frequency domains and others contribute additional task diversity grounded in everyday workflows. To ensure coverage beyond standardized interfaces, our benchmark includes numerous low-frequency domains (each $<$ 2\%), comprising legacy systems, departmental portals, and domain-specific dashboards. Collectively, these long-tail domains ($\simeq$ 15.5\%) are essential for evaluating agent generalization and robustness under web interface heterogeneity.

Tasks are grouped into \emph{four} categories based on required reasoning and interaction complexity. (1) \textbf{Data Extraction} tasks involve retrieving factual information from structured pages. (2) \textbf{Complex Reasoning} tasks require multi-hop inference, such as analyzing hierarchical structures. (3) \textbf{Table QnA} targets tabular interfaces, ranging from simple lookups to multi-dimensional aggregation. (4) \textbf{Form Filling} assesses the agent’s ability to translate natural language commands into precise UI actions, such as form submissions and event creation. This taxonomy enables a comprehensive evaluation of both the reasoning and execution capabilities of web automation agents across realistic enterprise scenarios.

\subsection{Demonstration Learning}
\label{subsec:demonstration_learning}
Website users are uniquely positioned to understand the optimal sequence of steps $S = {s_1, s_2, ..., s_n}$ required to complete online tasks. Their practical expertise is invaluable in creating Standard Operating Procedures (SOPs) for Cybernaut's execution. While a straightforward approach would involve users writing detailed textual prompts $P$ to describe the task, this method introduces certain limitations despite its simplicity and accessibility. Users often struggle to determine the appropriate level of detail, and their implicit assumptions or cognitive biases may lead to the omission of critical intermediate steps — steps that are essential for the reliable execution of tasks by LLMs.

To address these limitations, we propose a novel methodology that utilizes LLMs to analyze user-provided task demonstrations. Let $D = (T, E)$ represent a demonstration, where $T$ is the task definition and $E = {e_1, e_2, ..., e_n}$ is the execution trace containing a sequence of user actions. In this approach, users manually follow their SOP to complete a task and record the sequence of actions taken - herein referred to as the \emph{task execution trace} - for a given input. Alongside this trace, users also provide a high-level task definition that articulates the intended objective or outcome of the task. The LLM processes this demonstration to generate a generalizable SOP template $G(D)$ with placeholder variables $V = {v_1, v_2, ..., v_k}$. For each new task instance $i$, the model populates the placeholders with relevant contextual data $C_i$ before initiating execution, resulting in a concrete execution plan $E_i = G(D, C_i)$. 

Our approach leverages a custom browser extension that captures the details of web interaction in JSON format with an option to also enable audio and video recording. Although this format supports direct replay, it is brittle in practice — sensitive to dynamic changes in web page structure and unable to generalize across different inputs. To overcome these limitations, we process the recorded JSON through an LLM to generate robust, step-by-step SOP instructions. The transformation function $f: JSON \rightarrow SOP$ is implemented through our carefully designed prompting methodology, detailed in appendix~\ref{prompt template SOP}. At present, our framework supports only single demonstrations with linear browsing (non-branching sequential navigation through a website) task executions where $|E| = n$ for some finite $n$.

A crucial avenue for our future research is the expansion of our framework to incorporate audio and video demonstration data, alongside branched and conditional traces. In this enhanced model, the agent navigation actions will be represented as embeddings to compare execution videos of task instances. The execution paths will be represented as a directed acyclic graph, $G = (V, E)$. Finally, these representations will be fused to enable a more sophisticated and nuanced analysis of user demonstrations, capturing the intricacies of non-linear interactions and conditional behaviors.

\begin{figure}[!htbp]
\centering
\includegraphics[width=0.6\textwidth]{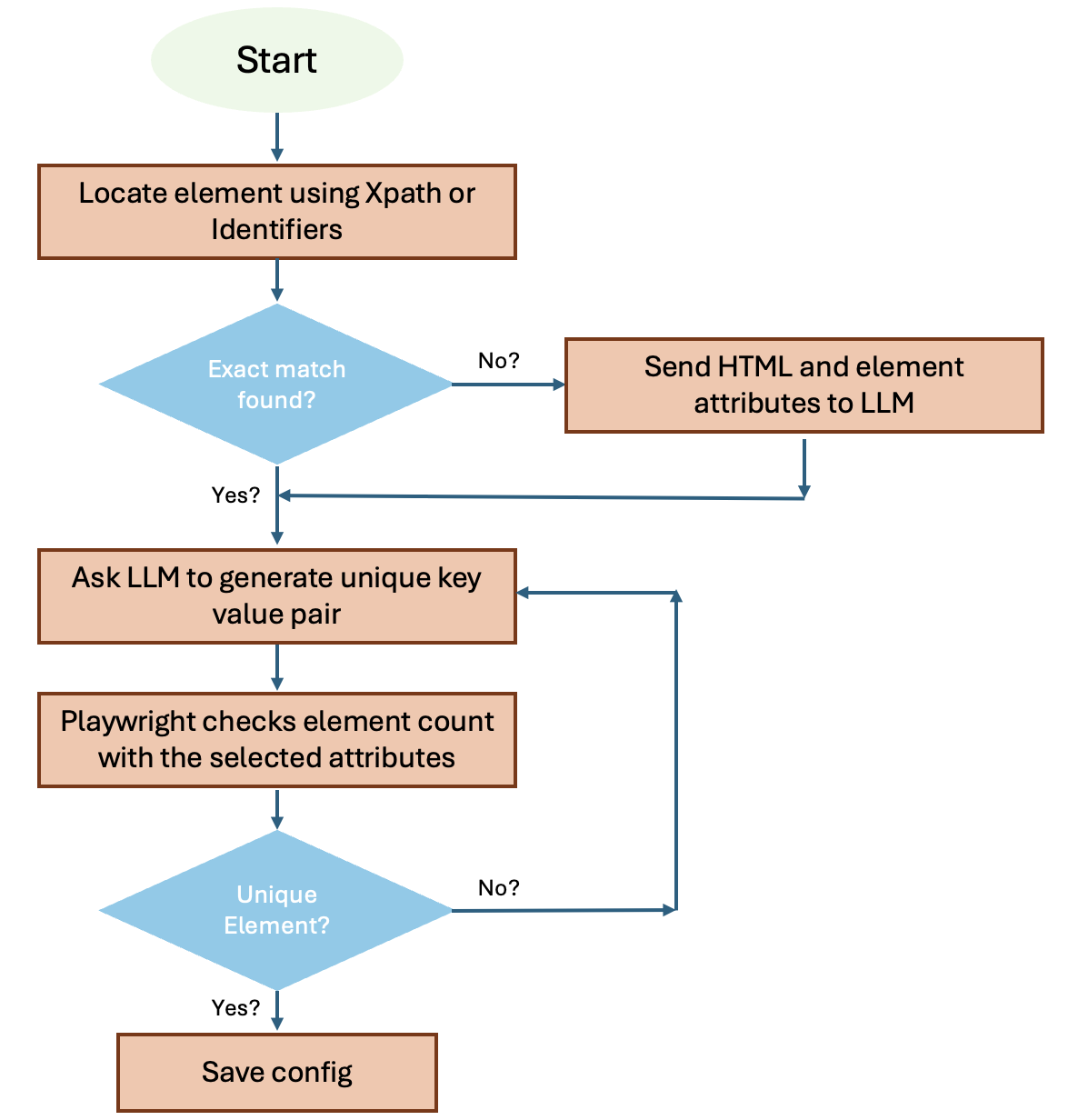}
\caption{Critical element identification approach}
\label{fig:element_identification}
\end{figure}

\subsection{Critical Element Identification}
\label{subsec:critical_element_identification}
A critical component of Cybernaut's functionality is the generation of an HTML snapshot that accurately identifies all interactive elements on a web page. The identification process relies on matching HTML element roles or tags and applying visibility filters. However, this approach encounters significant limitations when attempting to detect elements within complex web structures, particularly when the target element is obscured by multiple layers of HTML code. A common challenge arises when essential elements are rendered with zero dimensions or are visually hidden due to the prioritization of presentational elements. For instance, input text boxes are frequently overlaid with decorative div and span elements for styling purposes, effectively masking their presence in the DOM hierarchy. This layering technique, while beneficial for achieving desired visual effects, can interfere with automated detection mechanisms. Modern web applications often employ sophisticated CSS and JavaScript frameworks that further complicate this issue by dynamically manipulating element visibility and positioning, making traditional detection methods less reliable.

To overcome these limitations, we developed an approach that leverages JSON data from the demonstration data. We use HTML element attributes and XPath DOM identifiers, to generate task-specific configurations. These configurations are designed to maintain element visibility and interactability, regardless of their initial rendering state in the application. However, using XPath and identifiers directly presents two significant challenges. First, XPath identifiers and element selectors are frequently generated dynamically, resulting in inconsistent values across different sessions. Second, there exists a fundamental mismatch between demonstration and execution phases: during demonstration, users typically interact with visible, parent elements in the interface, while the actual functional elements are often hidden child components.

To systematically identify and enable interactive elements, we employ a web agent to execute the task. At each step, the agent applies a three-stage procedure for element identification, as shown in Figure~\ref{fig:element_identification}:

\textbf{1. Presence Verification:} At current page, we attempt to locate element using XPath and identifiers. If no exact match is found, this means the state has changed and XPaths are not valid anymore. To get the HTML code of element, we submit both the recorded element attributes and the current HTML snapshot to an LLM to perform semantic matching and return HTML code of element that might match this element.

\textbf{2. Key-Value Signature Assignment:} Upon a successful match, now we need to define some unique attributes of the element that we can save as configuration. To do that we employ an LLM to extract a set of stable key-value attribute pairs that uniquely identify the element. Once the attributes are identified, we use playwright to check how many element exists on this page with these attributes. If unique, we save it otherwise LLM is asked to generate new attributes.

\textbf{3. Configuration Persistence:} Finally, we store validated element signatures in a persistent configuration file that is read during execution. This config ensures consistent and automatic visibility toggling in future executions of the same task.

This method enhances element robustness and reduces fragility in web automation, particularly in dynamic or conditionally rendered user interfaces. For example, for a search input element from IMDb's interface:

\begin{verbatim}
{
    "tag": "input",
    "attributes": {
        "type": "text",
        "aria-label": "Search IMDb",
        "class": "imdb-header-search__input",
        "id": "suggestion-search",
        "data-testid": "suggestion-search"
    }
}
\end{verbatim}

Our system analyzes these attributes and generates a minimal yet robust configuration signature:

\begin{verbatim}
{
    "data-testid": "suggestion-search",
    "aria-label": "Search IMDb"
}
\end{verbatim}

\subsection{Task Execution Consistency}
\label{subsec:task_execution_consistency}
We define \emph{consistency} as Cybernaut's ability to reproduce similar execution patterns when performing identical tasks with same or varying input parameters. Inconsistent behavior emerges from three key factors: (1) LLM output stochasticity ($\sigma$), (2) dynamic form dependencies where $state_{t+1} = f(state_t, action_t)$, and (3) temporal website changes. While existing solutions employ deterministic selectors, prompt engineering, or retry mechanisms, they either sacrifice flexibility for consistency or impose computational overhead. More importantly, they fail to capture the fundamental relationship between successful execution patterns and intended task objectives.
 
Execution consistency is crucial to optimize Knowledge Operations Work (KOW) with well-defined tasks, as it enables human-level accuracy and streamlines debugging of automated workflows. To address this, we propose a trace-based similarity metric $C(E_t, E_g)$ that measures the alignment between an agent's execution path $E_t$ and a set of reference (golden) traces $E_g$, enabling systematic validation of task completion patterns and effective intervention during execution.
 
\subsubsection{Definition of Consistency}
\label{subsubsec:consistency_definition}
In web automation, \emph{consistency} refers to the similarity of execution patterns when performing the same task with identical or varying input parameters. We formalize this concept as follows:

Given a task $T$ and two input parameter sets $P_1$ and $P_2$, let $E_1$ and $E_2$ denote the execution traces corresponding to $T(P_1)$ and $T(P_2)$ respectively. The \emph{consistency score} $C$ between the two executions is defined as:
\begin{equation}
    \label{eq:consistency_score}
    C(E_1, E_2) = S(f(E_1), f(E_2))
\end{equation}
Here, $f$ is a feature extraction function, and $S$ is a similarity metric normalized to the interval $[0,1]$. A score of 1 indicates perfect consistency.

We decompose each execution trace into its constituent steps. For an execution trace $E$ consisting of $n$ steps $[e_1, e_2, \dots, e_n]$, each step $e_i$ is represented by a feature vector $h(e_i)$, defined as:
\begin{equation}
    \label{eq:feature_vector}
    h(e_i) = [g_i, a_i, \alpha_i]
\end{equation}
where, $g_i \in G$ denotes Cybernaut's goal state at step $i$, $a_i \in A$ denotes the action executed at step $i$, and, $\alpha_i \in \mathbb{R}^d$ is a $d$-dimensional attribute vector representing the properties of the web element interacted with at step $i$. The complete feature representation of an execution trace $E$ is then the ordered sequence of its step-level feature vectors: $f(E) = [h(e_1), h(e_2), \dots, h(e_n)]$. This formalization provides a principled foundation for quantitatively comparing execution behaviors and evaluating the consistency of web automation agents. 

\subsubsection{Calculating Consistency using LLM}
\label{subsubsec:consistency_calculation_llm}
Measuring consistency between execution traces presents unique challenges due to the inherent variability in LLM-driven automation. Even when executing identical tasks, the resulting action sequences may differ superficially while still achieving the same functional outcome. For example, consider two executions of the same form-filling task: \textbf{Execution 1:} [click input field $\rightarrow$ type text], and, \textbf{Execution 2:} [directly type text into input field]. Although these sequences produce different feature vectors $f(E_i)$, they are functionally equivalent. Similar variations may arise in scrolling behavior, element selection strategies, and interaction sequences. A robust consistency measure must account for these acceptable variations while detecting true deviations.

A straightforward approach to handle such nuanced comparisons is to leverage LLMs for consistency evaluation. Given two execution traces, an LLM can be prompted to assess their similarity based on the semantics of the actions, rather than exact structural alignment: $C(E_1, E_2) = LLM(f(E_1), f(E_2)) \rightarrow [0,1]$. This approach benefits from the LLM's natural language understanding capabilities, enabling flexible and semantic comparison without the need for manually engineered rules. It is particularly effective at recognizing variations that preserve task intent, such as reordering of steps or alternative interaction modalities. However, LLM-based evaluation is computationally expensive and may become infeasible at industrial scale, where millions of execution traces are analyzed. Moreover, it is susceptible to non-determinism — repeated evaluations on the same input pair can yield inconsistent scores. Achieving reliable results requires careful prompt engineering and strict control over evaluation conditions, which adds complexity and reduces reproducibility.

\subsubsection{Calculating Consistency using Embedding Models}
\label{subsubsec:consistency_calculation_embedding}
An alternative approach to LLM-based evaluation is the use of \emph{embedding models}, which strike a balance between semantic understanding and computational efficiency. These models encode execution traces into dense vector representations, enabling rapid and scalable similarity comparisons. Given two execution traces $E_1$ and $E_2$, the consistency score is defined as: 

\begin{equation}
C(E_1, E_2) = \text{Sim}(\text{Embed}(f(E_1)), \text{Embed}(f(E_2)))
\end{equation}
where $\text{Embed}(\cdot)$ is an embedding function and $\text{Sim}(\cdot,\cdot)$ is a similarity metric such as cosine similarity. 

These embedding-based methods offer a promising middle ground between the semantic richness of LLM-based evaluation and the efficiency requirements of real-world systems. They produce deterministic outputs, significantly reduce computational overhead, and support millisecond-level inference times—making them well-suited for high-throughput environments. Once trained, embedding models provide consistent and reproducible evaluation criteria across millions of comparisons. Given these advantages, we adopt the embedding-based approach as our final solution for consistency computation in the proposed workflow.

\section{Performance Evaluation}
\label{sec:performance_evaluation}

\subsection{Experimental Setup}
\label{subsec:experimental_setup}
\textbf{SOP Generation from User Demonstration:} We leveraged custom browser extension to capture web interaction sequences in JSON format. The recorded demonstrations along with task definitions were then processed by an LLM (Clause 3.7) to generate structured SOPs.

\textbf{Embedding Model Data Collection and Preparation:} To fine-tune the consistency embedding model, we constructed a labeled dataset by executing each task 10 times using Cybernaut. We then manually reviewed and paired the resulting execution traces, annotating each pair as either \emph{similar} or \emph{dissimilar}. The fine-tuning dataset includes 494 similar and 322 dissimilar pairs, derived from 80 executions across 8 distinct task types. Labeling was based on the principle of functional equivalence rather than exact action sequence matching. Non-critical variations - such as additional scrolls, redundant clicks, or differences in timing, were deliberately ignored. Instead, our annotations prioritized the "spirit of execution", aiming to capture whether the agent achieved the same outcome through a logically coherent and consistent behavior pattern. 


\textbf{Embedding Model Training using Siamese Network with Contrastive Loss:} We employed a Siamese network architecture to fine-tune the consistency embedding model. Each execution trace pair was processed through identical network branches that produced fixed-length 768-dimensional embedding via mean pooling over step-level feature vectors. The network was trained using a contrastive loss function, which minimized the distance between embedding of similar traces while maximizing it for dissimilar ones. Fine-tuning was performed using \textit{all-distilroberta-v1} over 816 labeled pairs for 3 epochs, using a learning rate of 5e-5 and a weight decay of 0.01. We used a binary classification evaluator for validation. This approach enabled the model to capture semantically meaningful patterns in execution behavior, forming a robust basis for consistency scoring.

\textbf{Cybernaut Agent Setup:} We extended the \texttt{Browser-Use} framework (version 0.1.40) to support task reproducibility and consistency monitoring. However, our solution can be extended to any web agent. Enhancements include (1) configuration support for saving interacted element signatures during onboarding, (2) customized tool invocations, (3) improved answer validation, and (4) robust URL handling. All LLM-driven operations inside cybernaut were powered by Claude 3.7, ensuring uniform behavior across planning and execution components.

\subsection{Results}
\label{subsec:results}

\subsubsection{Task Completion Evaluation}
\label{subsec:task_completion_evaluation}
In this section, we present the evaluation results of Cybernaut on our company's internal benchmark, highlighting the impact of demonstration-based SOP generation and critical element handling on task completion accuracy. For the public WebVoyager benchmark \cite{he-etal-2024-webvoyager}, we did not generate demonstrations, as most tasks lack clearly defined ground truth and reproducible step sequences. Nevertheless, for completeness, we evaluate Cybernaut (configured without SOPs but with critical element handling) on this benchmark and compare its performance against the Browser-Use baseline. Cybernaut achieves comparable accuracy (80.3\% v/s 82.2\% for SOTA) on the WebVoyager benchmark, despite the absence of demonstrations. For further details, please refer to section~\ref{webvoyager-benchmark}.

Table~\ref{tab:cybernaut_internal} summarizes the accuracy improvements on the internal benchmark dataset. The integration of SOP alone yields a 13.9\% accuracy improvement (from 72\% to 82.02\%) over the state-of-the-art (SOTA) baseline \texttt{Browser-Use}. Further incorporating the critical element detection and handling module leads to an additional 9.3\% gain over SOTA, resulting in a final accuracy of 88.68\%.

\begin{table}[!htbp]
  \caption{Cybernaut accuracy on internal benchmark}
  \centering
  \begin{tabular}{lll}
    \toprule
    Model & Accuracy \\
    \midrule
    Browser-Use & 72.00\% \\
    Cybernaut with SOP & 82.02\% \\
    \textbf{Cybernaut with SOP + Critical element fix} & \textbf{88.68\%} \\
    \bottomrule
  \end{tabular}
  \label{tab:cybernaut_internal}
\end{table}



\subsubsection{Task Consistency Evaluation}
\label{subsec:consistency_evaluation}
Table~\ref{tab:consistency_evaluation} presents the results of the consistency model evaluated on a manually labeled validation dataset. The out-of-the-box model performs poorly, requiring an impractically high prediction threshold of 99.8\% to achieve 71.1\% accuracy and 72.6\% F1 score, due to its lack of prior exposure to structured web execution traces. In contrast, after fine-tuning, the Siamese model achieves superior performance with 84.7\% accuracy and 87.3\% F1 score at a more reasonable threshold of 81.1\%, demonstrating its enhanced capability to differentiate between consistent and inconsistent execution patterns.

\begin{table}[!htbp]
  \caption{Consistency model performance on validation set}
  \centering
  \begin{tabular}{lccc}
    \toprule
    Model Type & Prediction threshold & Accuracy & F1 Score\\
    \midrule
    Out-of-the-box &99.8\% & 71.1\% & 72.6\% \\
    Fine-tuned & 81.1\% &84.7\% & 87.3\% \\
    \bottomrule
  \end{tabular}
  \label{tab:consistency_evaluation}
\end{table}

\begin{table}[!htbp]
\caption{Execution consistency analysis: similarity score comparison}
\centering
\begin{tabular}{lc}
\toprule
Execution Type & Average Similarity Score (\%) \\
\midrule
Consistent & 90.97 \\
Inconsistent & 60.60 \\
\bottomrule
\end{tabular}
\label{tab:similar_and_dissimilar_tasks_score_diff} \\
\textit{Note}: Similarity scores computed using cosine similarity between execution traces. Consistent executions demonstrate significantly higher similarity scores, indicating reliable task reproduction.
\end{table}

During run-time, the consistency model will compare execution traces with reference traces to assess similarity. Analysis of cosine similarities between known executions shows consistent cases achieving 90.97\% similarity versus 60.60\% for inconsistent ones as shown in Table~\ref{tab:similar_and_dissimilar_tasks_score_diff}. This margin demonstrates the model's ability to differentiate equivalent executions, validating its use in system monitoring.

\subsection{Performance on WebVoyager Public Benchmark Dataset}
\label{webvoyager-benchmark}
For the WebVoyager benchmark dataset (643 tasks across 15
websites) \cite{he-etal-2024-webvoyager}, we did not generate demonstrations, as most tasks lack a clearly defined ground truth and reproducible step sequences. In many cases, the clicked links vary based on external factors such as product updates (e.g., change in airpods version) and dynamic recommendation changes for the task (e.g., different headphones recommendation). Nevertheless, for completeness, we evaluate Cybernaut on this benchmark and compare it with the Browser-Use baseline. As shown in Table~\ref{tab:cybernaut_webvoyager}, Cybernaut achieves an accuracy of 80.3\%, representing a marginal 2.37\% drop compared to the SOTA baseline. A detailed analysis reveals that this discrepancy is not concentrated in any specific domain. In fact, several tasks are successfully completed by Cybernaut but not by Browser-Use, and vice versa. The variations are primarily attributable to non-determinism in LLM behavior, \texttt{CAPTCHA} blocks on external websites, and dynamic content differences.

\begin{table}[!htbp]
  \caption{Cybernaut accuracy on Webvoyager benchmark}
  \centering
  \begin{tabular}{lll}
    \toprule
    Model & Accuracy \\
    \midrule
    \textbf{Browser-Use} & \textbf{82.20\%} \\
    Cybernaut & 80.30\% \\
    \bottomrule
  \end{tabular}
  \label{tab:cybernaut_webvoyager}
\end{table}

\section{Conclusion and Future Work}
\label{sec:conclusion}
In this paper, we presented \emph{Cybernaut}, an advanced framework for web automation AI agent designed to address critical challenges in enterprise environments. Empirical evaluation demonstrated an 88.68\% task completion rate on internal benchmarks, representing a 23.2\% improvement over baseline methods. Our consistency evaluation method, powered by a fine-tuned embedding model, effectively distinguishes between consistent and inconsistent execution patterns with 84.7\% accuracy, enabling reliability at industrial scale. These results establish Cybernaut as a robust and and generalizable solution for highly accurate and consistent enterprise-grade web automation, and provide a foundation for future advancements. 

Future work will explore \emph{multi-step demonstration learning} to handle more complex workflows involving conditional execution traces, as well as the integration of visual information (e.g., page screenshots) alongside user recorded JSON to improve element recognition accuracy. Additionally, we plan to investigate graph-based approaches for modeling execution path structures and transitions, and to enhance our consistency evaluation framework using visual context to better capture UI state dynamics. Also, we will extend consistency metrics to run during execution and nudge model in right direction if model deviates from previous consistent paths.




\begin{thebibliography}{10}

\bibitem{openai2025cua}
OpenAI.
\newblock Computer-Using Agent.
\newblock \url{https://openai.com/index/computer-using-agent/}, 2025.

\bibitem{anthropic2024computeruse}
Anthropic.
\newblock Computer Use (Beta).
\newblock \url{https://docs.anthropic.com/en/docs/agents-and-tools/computer-use}, October 2024.

\bibitem{browser_use2024}
Magnus Müller and Gregor Žunič.
\newblock Browser Use: Enable AI to control your browser.
\newblock GitHub, 2024.
\newblock \url{https://github.com/browser-use/browser-use}.

\bibitem{nova_act}
Amazon.
\newblock Nova Act.
\newblock \url{https://nova.amazon.com/act}, 2025.

\bibitem{lu_steve_stepverif_2025}
Fanbin Lu, Zhisheng Zhong, Ziqin Wei, Shu Liu, Chi-Wing Fu, and Jiaya Jia.
\newblock STEVE: A Step Verification Pipeline for Computer-use Agent Training.
\newblock {\em arXiv preprint arXiv:2503.12532}, 2025.

\bibitem{he-etal-2024-webvoyager}
Hongliang He, Wenlin Yao, Kaixin Ma, Wenhao Yu, Yong Dai, Hongming Zhang, Zhenzhong Lan, and Dong Yu.
\newblock WebVoyager: Building an End-to-End Web Agent with Large Multimodal Models.
\newblock In {\em Proceedings of the 62nd Annual Meeting of the Association for Computational Linguistics (Volume 1: Long Papers)}, pages 6864--6890, Bangkok, Thailand, August 2024. Association for Computational Linguistics.

\bibitem{deng2024_mid2web_neurips}
Xiang Deng, Yu Gu, Boyuan Zheng, Shijie Chen, Sam Stevens, Boshi Wang, Huan Sun, and Yu Su.
\newblock Mind2Web: Towards a Generalist Agent for the Web.
\newblock In {\em Advances in Neural Information Processing Systems}, volume 36, pages 28091--28114. Curran Associates, Inc., 2023.

\bibitem{koh2024_visualwebarena}
Jing Yu Koh, Robert Lo, Lawrence Jang, Vikram Duvvur, Ming Chong Lim, Po-Yu Huang, Graham Neubig, Shuyan Zhou, Ruslan Salakhutdinov, and Daniel Fried.
\newblock VisualWebArena: Evaluating Multimodal Agents on Realistic Visual Web Tasks.
\newblock {\em arXiv preprint arXiv:2401.13649}, 2024.

\bibitem{rawles2025_androidworld}
Christopher Rawles, Sarah Clinckemaillie, Yifan Chang, Jonathan Waltz, Gabrielle Lau, Marybeth Fair, Alice Li, William Bishop, Wei Li, Folawiyo Campbell-Ajala, Daniel Toyama, Robert Berry, Divya Tyamagundlu, Timothy Lillicrap, and Oriana Riva.
\newblock AndroidWorld: A Dynamic Benchmarking Environment for Autonomous Agents.
\newblock {\em arXiv preprint arXiv:2405.14573}, 2025.

\bibitem{xu2024_crabxenv_bench}
Tianqi Xu, Linyao Chen, Dai-Jie Wu, Yanjun Chen, Zecheng Zhang, Xiang Yao, Zhiqiang Xie, Yongchao Chen, Shilong Liu, Bochen Qian, Anjie Yang, Zhaoxuan Jin, Jianbo Deng, Philip Torr, Bernard Ghanem, and Guohao Li.
\newblock CRAB: Cross-environment Agent Benchmark for Multimodal Language Model Agents.
\newblock {\em arXiv preprint arXiv:2407.01511}, 2024.

\bibitem{song2025_api_browsingapi}
Yueqi Song, Frank Xu, Shuyan Zhou, and Graham Neubig.
\newblock Beyond Browsing: API-Based Web Agents.
\newblock {\em arXiv preprint arXiv:2410.16464}, 2025.

\bibitem{lu2024_weblinx}
Xing Han Lù, Zdeněk Kasner, and Siva Reddy.
\newblock WebLINX: Real-World Website Navigation with Multi-Turn Dialogue.
\newblock {\em arXiv preprint arXiv:2402.05930}, 2024.

\bibitem{zheng2024_gpt4_webagent}
Boyuan Zheng, Boyu Gou, Jihyung Kil, Huan Sun, and Yu Su.
\newblock GPT-4V(ision) is a Generalist Web Agent, if Grounded.
\newblock {\em arXiv preprint arXiv:2401.01614}, 2024.

\bibitem{cai2025_PUMA_webagent}
Hongru Cai, Yongqi Li, Wenjie Wang, Fengbin Zhu, Xiaoyu Shen, Wenjie Li, and Tat-Seng Chua.
\newblock Large Language Models Empowered Personalized Web Agents.
\newblock In {\em Proceedings of the ACM on Web Conference 2025}, pages 198--215. Association for Computing Machinery, 2025.

\bibitem{wang2025_guiagentsfoundation}
Shuai Wang, Weiwen Liu, Jingxuan Chen, Yuqi Zhou, Weinan Gan, Xingshan Zeng, Yuhan Che, Shuai Yu, Xinlong Hao, Kun Shao, Bin Wang, Chuhan Wu, Yasheng Wang, Ruiming Tang, and Jianye Hao.
\newblock GUI Agents with Foundation Models: A Comprehensive Survey.
\newblock {\em arXiv preprint arXiv:2411.04890}, 2025.

\bibitem{wen2024a_autodroid_automation}
Hao Wen, Yuanchun Li, Guohong Liu, Shanhui Zhao, Tao Yu, Toby Jia-Jun Li, Shiqi Jiang, Yunhao Liu, Yaqin Zhang, and Yunxin Liu.
\newblock AutoDroid: LLM-powered Task Automation in Android.
\newblock {\em arXiv preprint arXiv:2308.15272}, 2024.

\bibitem{wen2024b_droidbotgpt_automation}
Hao Wen, Hongming Wang, Jiaxuan Liu, and Yuanchun Li.
\newblock DroidBot-GPT: GPT-powered UI Automation for Android.
\newblock {\em arXiv preprint arXiv:2304.07061}, 2024.

\bibitem{li2020_mappinglang_instructions}
Yang Li, Jiacong He, Xin Zhou, Yuan Zhang, and Jason Baldridge.
\newblock Mapping Natural Language Instructions to Mobile UI Action Sequences.
\newblock {\em arXiv preprint arXiv:2005.03776}, 2020.

\bibitem{hong2024cogagentvisuallanguagemodel}
Wenyi Hong, Weihan Wang, Qingsong Lv, Jiazheng Xu, Wenmeng Yu, Junhui Ji, Yan Wang, Zihan Wang, Yuxuan Zhang, Juanzi Li, Bin Xu, Yuxiao Dong, Ming Ding, and Jie Tang.
\newblock CogAgent: A Visual Language Model for GUI Agents.
\newblock {\em arXiv preprint arXiv:2312.08914}, 2024.

\bibitem{zhan2025_appagent_acm}
Chi Zhang, Zhao Yang, Jiaxuan Liu, Yanda Li, Yucheng Han, Xin Chen, Zebiao Huang, Bin Fu, and Gang Yu.
\newblock AppAgent: Multimodal Agents as Smartphone Users.
\newblock In {\em Proceedings of the 2025 CHI Conference on Human Factors in Computing Systems}, Article 70, pages 1--20. Association for Computing Machinery, 2025.

\bibitem{wang2024_mobileagent_multimod}
Junyang Wang, Haiyang Xu, Jiabo Ye, Ming Yan, Weizhou Shen, Ji Zhang, Fei Huang, and Jitao Sang.
\newblock Mobile-Agent: Autonomous Multi-Modal Mobile Device Agent with Visual Perception.
\newblock {\em arXiv preprint arXiv:2401.16158}, 2024.

\bibitem{you2024_ferretui_ground}
Keen You, Haotian Zhang, Eldon Schoop, Floris Weers, Amanda Swearngin, Jeffrey Nichols, Yinfei Yang, and Zhe Gan.
\newblock Ferret-UI: Grounded Mobile UI Understanding with Multimodal LLMs.
\newblock {\em arXiv preprint arXiv:2404.05719}, 2024.

\bibitem{lu2024_omniparser_vision}
Yadong Lu, Jianwei Yang, Yelong Shen, and Ahmed Awadallah.
\newblock OmniParser for Pure Vision Based GUI Agent.
\newblock {\em arXiv preprint arXiv:2408.00203}, 2024.

\bibitem{ge2025_irisb_guicomplex}
Zhiqi Ge, Juncheng Li, Xinglei Pang, Minghe Gao, Kaihang Pan, Wang Lin, Hao Fei, Wenqiao Zhang, Siliang Tang, and Yueting Zhuang.
\newblock Iris: Breaking GUI Complexity with Adaptive Focus and Self-Refining.
\newblock {\em arXiv preprint arXiv:2412.10342}, 2025.

\bibitem{ning2025survey}
Liangbo Ning, Ziran Liang, Zhuohang Jiang, Haohao Qu, Yujuan Ding, Wenqi Fan, Xiao-yong Wei, Shanru Lin, Hui Liu, Philip S. Yu, and Qing Li.
\newblock A Survey of WebAgents: Towards Next-Generation AI Agents for Web Automation with Large Foundation Models.
\newblock {\em arXiv preprint arXiv:2503.23350}, 2025.

\bibitem{correia2023survey}
André Correia and Luís A. Alexandre.
\newblock A Survey of Demonstration Learning.
\newblock {\em arXiv preprint arXiv:2303.11191}, 2023.

\end{thebibliography}

\newpage
\appendix

\section{Appendix A: User demonstration to step-by-step SOP instructions}
\label{sec:appendix:demonstration_learning}
To convert user demonstrations into step-by-step SOP instructions, we utilize an LLM that takes as input a high-level task definition and a browser-recorded execution trace in JSON format. As shown in Listing ~\ref{lst:execution-trace}, users perform the task manually following the SOP, while their interactions are captured via Chrome’s recorder. The LLM analyzes this interaction trace along with the task objective (e.g., identifying reusable laptop models within a specified age range) and, using a prompt template ~\ref{lst:sop_prompt}, generates a generalizable, step-by-step SOP ~\ref{lst:sop}. The prompt guides the model to infer user intent, abstract relevant actions, and exclude irrelevant or redundant interactions, resulting in a clear and reusable instruction set. This methodology enables precise documentation of operational workflows, facilitating reliable automation by downstream web agents.

\begin{lstlisting}[language=json, caption={A sample execution trace from Recorder (JSON)}, label={lst:execution-trace}]
{
    "title": "Recording IMDB",
    "steps": [
        {
            "type": "setViewport",
            "width": 1173,
            "height": 901,
            "deviceScaleFactor": 1,
            "isMobile": false,
            "hasTouch": false,
            "isLandscape": false
        },
        {
            "type": "navigate",
            "url": "https://www.imdb.com/",
            "assertedEvents": [
                {
                    "type": "navigation",
                    "url": "https://www.imdb.com/",
                    "title": ""
                }
            ]
        },
        {
            "type": "click",
            "target": "main",
            "selectors": [
                [
                    "aria/All",
                    "aria/[role=\"none\"]"
                ],
                [
                    "span.sc-bBjSGg svg"
                ],
                [
                    "xpath///*[@data-testid=\"category-selector-button\"]/svg"
                ],
                [
                    "pierce/span.sc-bBjSGg svg"
                ]
            ],
            "offsetY": 11,
            "offsetX": 2.1796875
        },
        {
            "type": "click",
            "target": "main",
            "selectors": [
                [
                    "#suggestion-search-container a > span.ipc-list-item__text"
                ],
                [
                    "xpath///*[@id=\"nav-search-form\"]/div[1]/div/div/div/div/ul/a/span[1]"
                ],
                [
                    "pierce/#suggestion-search-container a > span.ipc-list-item__text"
                ],
                [
                    "text/Advanced Search"
                ]
            ],
            "offsetY": 12.359375,
            "offsetX": 97.8671875,
            "assertedEvents": [
                {
                    "type": "navigation",
                    "url": "https://www.imdb.com/search/title/?ref_=nv_sr_menu_adv",
                    "title": ""
                }
            ]
        },
        {
            "type": "click",
            "target": "main",
            "selectors": [
                [
                    "aria/Expand all",
                    "aria/[role=\"generic\"]"
                ],
                [
                    "div.sc-ed40b8bf-0 > div span"
                ],
                [
                    "xpath///*[@data-testid=\"adv-search-expand-all\"]/span"
                ],
                [
                    "pierce/div.sc-ed40b8bf-0 > div span"
                ],
                [
                    "text/Expand all"
                ]
            ],
            "offsetY": 11,
            "offsetX": 60.7578125
        },
        {
            "type": "click",
            "target": "main",
            "selectors": [
                [
                    "aria/Enter release date above[role=\"textbox\"]"
                ],
                [
                    "[data-testid='releaseYearMonth-start']"
                ],
                [
                    "xpath///*[@data-testid=\"releaseYearMonth-start\"]"
                ],
                [
                    "pierce/[data-testid='releaseYearMonth-start']"
                ]
            ],
            "offsetY": 35,
            "offsetX": 92.5
        },
        {
            "type": "change",
            "value": "2020-01",
            "selectors": [
                [
                    "aria/Enter release date above[role=\"textbox\"]"
                ],
                [
                    "[data-testid='releaseYearMonth-start']"
                ],
                [
                    "xpath///*[@data-testid=\"releaseYearMonth-start\"]"
                ],
                [
                    "pierce/[data-testid='releaseYearMonth-start']"
                ]
            ],
            "target": "main"
        },
        {
            "type": "click",
            "target": "main",
            "selectors": [
                [
                    "aria/Enter release date below[role=\"textbox\"]"
                ],
                [
                    "[data-testid='releaseYearMonth-end']"
                ],
                [
                    "xpath///*[@data-testid=\"releaseYearMonth-end\"]"
                ],
                [
                    "pierce/[data-testid='releaseYearMonth-end']"
                ]
            ],
            "offsetY": 27,
            "offsetX": 33.90625
        },
        {
            "type": "change",
            "value": "2020-12",
            "selectors": [
                [
                    "aria/Enter release date below[role=\"textbox\"]"
                ],
                [
                    "[data-testid='releaseYearMonth-end']"
                ],
                [
                    "xpath///*[@data-testid=\"releaseYearMonth-end\"]"
                ],
                [
                    "pierce/[data-testid='releaseYearMonth-end']"
                ]
            ],
            "target": "main"
        },
        {
            "type": "click",
            "target": "main",
            "selectors": [
                [
                    "[data-testid='autosuggest-input-test-id-languages']"
                ],
                [
                    "xpath///*[@data-testid=\"autosuggest-input-test-id-languages\"]"
                ],
                [
                    "pierce/[data-testid='autosuggest-input-test-id-languages']"
                ]
            ],
            "offsetY": 12,
            "offsetX": 116.5
        },
        {
            "type": "change",
            "value": "japa",
            "selectors": [
                [
                    "[data-testid='autosuggest-input-test-id-languages']"
                ],
                [
                    "xpath///*[@data-testid=\"autosuggest-input-test-id-languages\"]"
                ],
                [
                    "pierce/[data-testid='autosuggest-input-test-id-languages']"
                ]
            ],
            "target": "main"
        },
        {
            "type": "click",
            "target": "main",
            "selectors": [
                [
                    "aria/Japanese"
                ],
                [
                    "[data-testid='checkbox-test-id-ja']"
                ],
                [
                    "xpath///*[@data-testid=\"checkbox-test-id-ja\"]"
                ],
                [
                    "pierce/[data-testid='checkbox-test-id-ja']"
                ]
            ],
            "offsetY": 26,
            "offsetX": 28.5
        },
        {
            "type": "click",
            "target": "main",
            "selectors": [
                [
                    "aria/See results"
                ],
                [
                    "[data-testid='adv-search-get-results']"
                ],
                [
                    "xpath///*[@data-testid=\"adv-search-get-results\"]"
                ],
                [
                    "pierce/[data-testid='adv-search-get-results']"
                ]
            ],
            "offsetY": 29,
            "offsetX": 60.5
        },
        {
            "type": "click",
            "target": "main",
            "selectors": [
                [
                    "aria/Sort by"
                ],
                [
                    "#adv-srch-sort-by"
                ],
                [
                    "xpath///*[@id=\"adv-srch-sort-by\"]"
                ],
                [
                    "pierce/#adv-srch-sort-by"
                ],
                [
                    "text/POPULARITY"
                ]
            ],
            "offsetY": 17.2734375,
            "offsetX": 61.3515625
        },
        {
            "type": "change",
            "value": "USER_RATING_COUNT",
            "selectors": [
                [
                    "aria/Sort by"
                ],
                [
                    "#adv-srch-sort-by"
                ],
                [
                    "xpath///*[@id=\"adv-srch-sort-by\"]"
                ],
                [
                    "pierce/#adv-srch-sort-by"
                ],
                [
                    "text/POPULARITY"
                ]
            ],
            "target": "main"
        },
        {
            "type": "click",
            "target": "main",
            "selectors": [
                [
                    "aria/Ascending sort order",
                    "aria/[role=\"none\"]"
                ],
                [
                    "[data-testid='test-sort-order'] > svg"
                ],
                [
                    "xpath///*[@data-testid=\"test-sort-order\"]/svg"
                ],
                [
                    "pierce/[data-testid='test-sort-order'] > svg"
                ]
            ],
            "offsetY": 15,
            "offsetX": 9.1796875
        }
    ]
}

\end{lstlisting}

\section{Appendix B: Prompt Template to generate SOP from user demonstration}
\label{prompt template SOP}
\begin{lstlisting}[language=yaml, caption={Prompt Template (YAML) for converting User Demonstration to SOP}, breaklines=true, breakindent=0pt, columns=flexible, label={lst:sop_prompt}]
prompts:  
  generator: |
    <role>
    You are a professional operations manager whose expertise is to document Standard Operating Procedure (SOP) in a clear and precise manner.
    This SOP will be used as a step-by-step instruction for an AI-powered browsing agent to complete similar tasks.
    </role>

    You are now given a demo of the operation procedure performed by a human associate for the following task described within <task_description> tags:
    <task_description>
    <INPUT_TASK_DESCRIPTION_EXAMPLE>
    </task_description>

    The demo peration procedure is recorded as a browser replay in .json format within the <browser_replay_in_json> tags:
    <demo>
    <browser_replay_in_json>
    <TEXT_REPLAY>
    </browser_replay_in_json>
    </demo>
    
    You are asked to compose an SOP which an AI-powered browsing agent can follow and complete similar tasks within the <task_for_sop> tags below:
    <task_for_sop>
    <INPUT_TASK_DESCRIPTION_GENERAL>
    </task_for_sop>
    
    Below is the requirement of what should be included in the SOP that you are going to compose:
    <requirement>
    1. For the first step in your SOP:
      - Use both website name and the exact URL in your instruction
    2. For the second step and onwards in your SOP:
      - Look holistically at the demo and identify the intention and goal behind each browsing action and step recorded. 
      - Use the intention and purpose behind to guide your composition of SOP. 
      - If navigating to a specific website is a critical action to achieve the goal, always include the web page name and the exact URL of this navigatio action.
      - For other actions, e.g., mouse clicks, keyboard inputs, that are related to the task, include them as illustration examples in your instrcution.
    3. Your SOP should only include the knowledge that can be drawn from the demo replay within <demo> tags. 
      - Do not come up with an SOP from your memory.
    4. Exclude steps within <demo> tages that are unrelated to the task within <task_for_sop> tags. Examples of such steps are:
      - Steps related to solving CAPTCHA.
      - Steps related to close pop up windows.
      - Mouse clicks on non-interactable elements such as background, or plain text.
    </requirement>

    You should follow the formatting instructions below to provide an SOP as your final answer:
    <format_instructions>
    1. Using <sop> tags to include all contents below.
    2. Restate the task description content within <task_for_sop> tags above, now using <task> tags.
    3. <task_for_sop> may be a general version of the <demo> example. Please identify proper input parameters so that <task_for_sop> can be faithfully represented.
      - Format the input parameters as .json within <input_param> tags. E.g., {input_param_1: "Explanation"}.
      - If no input parameter needed, output an empty dict within <input_param> tags. I.e., {}.
    4. Document your final SOP within <instructions-step-by-step> tags.
      - Only provide the instructions within <instructions-step-by-step> tags. No need to explain within <instructions-step-by-step> tags.
      - Use proper referece to the input parameters identified. E.g., using <INPUT_PARAMETER_1>
      - Using numbered points to organize your sop.
    </format_instructions>

  guardrail: |
    Additional guardrails that you are asked to follow:
    <guardrails>
    1. Only navigate the websites you are in. Do not open any URL from your memory.
    </guardrails>
  
  output_format: |
    Below are the output formatting requirements:
    <output_format>
    1. Provide your final answer to the task within <final_answer> tags. 
      - If you cannot find the answer related to the <task> from your browsing activities, say "No answer found". Do not make up one yourself based on your prior knowledge.
    2. Provide your thoughts within <thoughts> tags.
    </output_format>
  
  prompt_head: |
    You are a browsing agent and are asked to perform a web browsing task described within <task> tags below.

    You are ask to act like a human associate who will strictly follow the instructions within <instructions-step-by-step> tags, using the appropriate input parameter values provided within the <input_param> tags.
\end{lstlisting}

\newpage
\section{Appendix C: Sample SOP generated from the prompt}
\begin{lstlisting}[language=yaml, caption={User Demonstration translated to a sample robust SOP}, breaklines=true, breakindent=0pt, columns=flexible, label={lst:sop}]

<sop>
<task>
Navigate to https://www.imdb.com/?ref_=nv_home and give me top 20 movies and shows in Japanese language with highest number of ratings in year 2020.
</task>

<input_param>
{
   "language": "Japanese",
  "year": "2020",
  "sort_by": "USER_RATING_COUNT"
}
</input_param>

<instructions-step-by-step>
1. Navigate to https://www.imdb.com/?ref_=nv_home
2. Click on the "All" dropdown menu at the top of the page (usually located in the search bar area)
3. Click on "Advanced Search" from the dropdown menu
4. In the Advanced Search page, first click on Expand all to access all search filters and then click on "Expand Release date" section
5. Enter "<year>-01" and "<year>-12" in the "Enter release date above" field
6. In the language search field, type "<language>" (e.g., "Japanese")
7. Select the checkbox for "<language>" from the dropdown results
8. Click on the "See results" button
9. On the results page, click on the "Sort by" dropdown menu
10. Select "<sort_by>" from the dropdown options (e.g., "USER_RATING_COUNT")
11. If needed, click on the sort order button to change from ascending to descending order
12. The page will now display the top 20 movies from <country> in <language> language with the highest number of ratings in <year>
</instructions-step-by-step>
</sop>
\end{lstlisting}

\end{document}